\documentclass[sigconf]{acmart}
\AtBeginDocument{%
  \providecommand\BibTeX{{%
    \normalfont B\kern-0.5em{\scshape i\kern-0.25em b}\kern-0.8em\TeX}}}

\copyrightyear{2024}
\acmYear{2024}
\setcopyright{rightsretained}

\acmConference[UIST Adjunct '24]{The 37th Annual ACM Symposium on User Interface Software and Technology}{October 13--16, 2024}{Pittsburgh, PA, USA}

\acmBooktitle{The 37th Annual ACM Symposium on User Interface Software and Technology (UIST Adjunct '24), October 13--16, 2024, Pittsburgh, PA, USA}
\acmDOI{10.1145/3672539.3686738}
\acmISBN{979-8-4007-0718-6/24/10}

\usepackage{nawta}

\begin{document}

\title{Real-Time Word-Level Temporal Segmentation in Streaming Speech Recognition}

\author{Naoto Nishida}
\email{nawta@g.ecc.u-tokyo.ac.jp}
\orcid{0000-0001-9966-4664}
\affiliation{%
  \institution{The University of Tokyo}
  \city{Tokyo}
  \country{Japan}
}

\author{Hirotaka Hiraki}
\email{hirotakahiraki@gmail.com}
\orcid{0000-0002-6543-4593}
\affiliation{%
  \institution{The University of Tokyo}
  \city{Tokyo}
  \country{Japan}
}
\affiliation{%
  \institution{National Institute of Advanced Industrial Science and Technology}
  \city{Chiba}
  \country{Japan}
}

\author{Jun Rekimoto}
\email{rekimoto@acm.org}
\orcid{0000-0002-3629-2514}
\affiliation{%
  \institution{The University of Tokyo}
  \city{Tokyo}
  \country{Japan}
}
\affiliation{%
  \institution{Sony CSL Kyoto}
  \city{Kyoto}
  \country{Japan}
}

\author{Yoshio Ishiguro}
\orcid{0000-0002-1781-6212}
\email{ishiy@acm.org}
\affiliation{%
  \institution{The University of Tokyo}
  \city{Tokyo}
  \country{Japan}
}


\renewcommand{\shortauthors}{Nishida, et al.}

\begin{abstract}

Rich-text captions are essential to help communication for Deaf and hard-of-hearing (DHH) people, second-language learners, and those with autism spectrum disorder (ASD). They also preserve nuances when converting speech to text, enhancing the realism of presentation scripts and conversation or speech logs. 
However, current real-time captioning systems lack the capability to alter text attributes (ex. capitalization, sizes, and fonts) at the word level, hindering the accurate conveyance of speaker intent that is expressed in the tones or intonations of the speech. 
For example, ``YOU should do this'' tends to be considered as indicating ``You'' as the focus of the sentence, whereas ``You should do THIS'' tends to be ``This'' as the focus.
This paper proposes a solution that changes the text decorations at the word level in real time. 
As a prototype, we developed an application that adjusts word size based on the loudness of each spoken word. 
Feedback from users implies that this system helped to convey the speaker's intent, offering a more engaging and accessible captioning experience.
\end{abstract}



\begin{teaserfigure}
  \includegraphics[width=\textwidth]{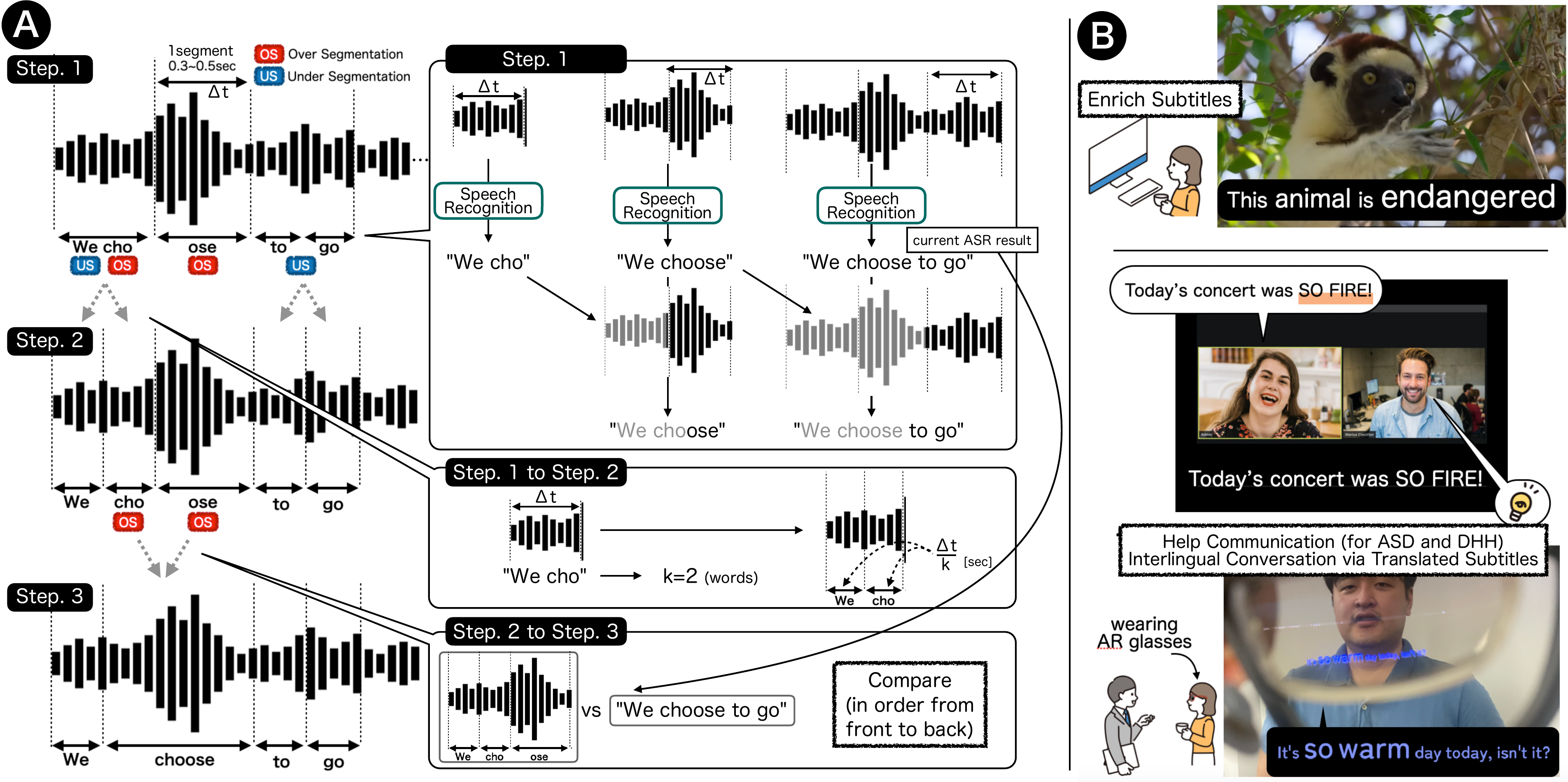}
  \caption{System Workflow and application examples. A) Our system temporally segments streaming Audio Speech Recognition (ASR) results in real time at word level. Workflow in detail is explained in Section \ref{sec:system_overview}.  B) Our system can be used to add word-level nonverbal information (e.g., volume, tones, and intonations) to texts such as those used for subtitles (upper), and to help communicate with people who find it hard to understand the nuances or people who are communicating via translated texts (middle and lower).}
  \Description{}
  \label{fig:teaser}
\end{teaserfigure}


\maketitle

\setlength\textfloatsep{2truemm}

\section{Introduction}







Rich text captions promise to significantly enhance communication for Deaf and hard-of-hearing (DHH) individuals~\cite{10.1145/3544548.3581130, 10.1145/3613904.3642258}, second language learners~\cite{10.1145/3613904.3642351}, and those with autism spectrum disorder (ASD)~\cite{10.1145/3613904.3642177}. By preserving nuances when converting speech to text, these captions can also increase the realism of presentation scripts and conversation logs, making them more engaging and effective. In addition, rich text captions can improve the entertainment value of movies and TV shows, creating a more immersive experience for viewers. The integration of advanced speech-to-text technologies into real-time captioning systems impacts how we communicate and consume media.

However, the problem is that current rich text captioning systems cannot alter text attributes (e.g., capitalizations, sizes, or fonts) at the word level in real-time. This limitation hinders the accurate conveyance of speaker intent, as nuances such as emphasis and intonation are not properly captured. For example, "\textit{YOU} should do this." versus "You should do \textit{THIS}." illustrates how varying emphasis can change the meaning.

To solve this problem, previous research has focused on developing real-time captioning systems that translate speech into text, but often overlook the importance of capturing and displaying paralinguistic cues. 
Some systems have attempted to use static text attributes or manual annotations, but these approaches are either not dynamic or require extensive manual input, making them impractical for real-time applications~\cite{10.1145/3544548.3581130, 10.1145/3544548.3581511}. Other tools such as whisper series~\cite{radford2022robustspeechrecognitionlargescale, whispertimestamped, fasterwhisper} need to spend relatively longer time to recognize (e.g., a few seconds to 30 seconds) the speech if it is used alone.
There remains a gap in developing a system that can dynamically alter word-level text attributes based on real-time speech analysis.
To fill in this gap, we design and implement a real-time rich-text captioning system that performs word-level temporal segmentation and timestamping using streaming speech recognition. 
As shown in Fig.~\ref{fig:teaser}, our system takes an input speech stream, segments it into words, assigns timestamps to each word, and dynamically adjusts the text decorations based on the detected paralinguistic cues such as loudness and emphasis. Specifically, the system works in three steps: first, it segments the audio signal according to frame rates, which can vary; second, it segments these frames further based on spaces or morphological analysis; third, it compares these segments with the latest Audio Speech Recognition (ASR) results to extract accurate timestamps and recognized words.


The main contribution of this paper is the development of a real-time rich-text captioning system that enhances the conveyance of speaker intent through dynamic word-level text decorations. In contrast to prior work that relied on static or manually annotated captions, our approach leverages advanced speech recognition technologies to provide a more engaging and accessible communication experience for a diverse range of users.
\section{System Overview}
\label{sec:system_overview}
As in Step 1, our system segments the audio signal to recognize the speech (illustrated as dotted lines in Figure \ref{fig:teaser}). The interval of each segment varies depending on the computers' capacity and conditions. It then recognizes the speech in the segment, where some may be over-segmented (ex. `ose' in at step 1 in Fig~\ref{fig:teaser}), under-segmented (ex. `to go'), or both occur (ex. `We cho'). To enhance the recognition accuracy, the system uses concatenated audio segments from the beginning to the newest segment.
From step 1 to step 2, it segments under-segmented segments according to space (or the results from morphological analysis in languages without spacing). We linearly divided $\Delta t$ into $k$ segments, each of which consists of $\Delta t / k$ seconds.
Therefore, all segments have over-segmented segments or properly segmented segments in step 2.
From step 2 to step 3, it compares step 2's result with the newest ASR result, which has proper grammatical segmentation information (which is spacing) to concatenate over-segmented parts (ex. `cho' and `ose' at step 2 in Fig.~\ref{fig:teaser}). 
As a result, our system extracts the segments of audio signals, their corresponding recognized words, and their timestamps of when the speech started and ended, illustrated at step 3 in Fig.~\ref{fig:teaser}. 
As the practical implementation detail, we used Unity to display the decorated subtitles. Therefore, the program was coded in C\# on the Unity platform. We also used Azure Speech Recognition API~\cite{azure} for ASR tasks. 
\section{Discussion}


Although promising, several challenges that have emerged during the use require further improvement.
First, the current integration of the over-segmented portions (Step 2 to Step 3) operates with a time complexity of $O(n^2)$, which is inefficient. Future work should focus on optimizing this step to enhance computational efficiency.
Second, our method for handling the sub-segmentation by dividing segments based on the word count per time step is just one approach. There is an inevitable error. 
Exploring alternative algorithms could improve accuracy and performance towards this error.
Third, additional applications for the system should be explored, such as real-time translation, educational tools for second language learners, and communication aids for ASD individuals. User studies will help validate these applications. 
Valuable past works would also help us design how to integrate nonspeech information in audio signals into captions to build useful applications for these people in our future work~\cite{10.1145/3597638.3608398}.
Fourth, addressing the delay in real-time processing is crucial. Minimizing lag will ensure natural communication flow.
Fifth, we currently use fixed timestamps for speech segments. Predicting the average duration of the speech from initial words and backcalculating start times could improve timing accuracy.
Sixth, our current integration method depends on the lag between steps. If processing speeds increase, this method may become unstable. Developing a more robust integration method is necessary.
Lastly, for languages without word spacing (e.g., Japanese), additional segmentation (e.g., MeCab~\cite{Kudo2005MeCabY}) is needed after Step 1.
Further user studies will validate the application examples and provide feedback for further refinement. Addressing these challenges will enhance the efficiency, accuracy, and versatility of our system.


\begin{acks}
This work was supported by JST BOOST Grant JPMJBS2418, JST Moonshot R\&D Grant JPMJMS2012, JST CREST Grant JPMJCR17A3, and the commissioned research by NICT Japan Grant JPJ012368C02901.
\end{acks}

\bibliographystyle{ACM-Reference-Format}
\bibliography{main}


\begin{thebibliography}{11}


\ifx \showCODEN    \undefined \def \showCODEN     #1{\unskip}     \fi
\ifx \showDOI      \undefined \def \showDOI       #1{#1}\fi
\ifx \showISBNx    \undefined \def \showISBNx     #1{\unskip}     \fi
\ifx \showISBNxiii \undefined \def \showISBNxiii  #1{\unskip}     \fi
\ifx \showISSN     \undefined \def \showISSN      #1{\unskip}     \fi
\ifx \showLCCN     \undefined \def \showLCCN      #1{\unskip}     \fi
\ifx \shownote     \undefined \def \shownote      #1{#1}          \fi
\ifx \showarticletitle \undefined \def \showarticletitle #1{#1}   \fi
\ifx \showURL      \undefined \def \showURL       {\relax}        \fi
\providecommand\bibfield[2]{#2}
\providecommand\bibinfo[2]{#2}
\providecommand\natexlab[1]{#1}
\providecommand\showeprint[2][]{arXiv:#2}

\bibitem[Azure SpeechServices({[n.\,d.]})]%
        {azure}
Azure SpeechServices \bibinfo{year}{[n.\,d.]}\natexlab{}.
\newblock
\newblock
\newblock
\shownote{\url{https://azure.microsoft.com/en-us}}.


\bibitem[Bhatia et~al\mbox{.}(2024)]%
        {10.1145/3613904.3642351}
\bibfield{author}{\bibinfo{person}{Arpit Bhatia}, \bibinfo{person}{Henning
  Pohl}, \bibinfo{person}{Teresa Hirzle}, \bibinfo{person}{Hasti Seifi}, {and}
  \bibinfo{person}{Kasper Hornb\ae{}k}.} \bibinfo{year}{2024}\natexlab{}.
\newblock \showarticletitle{Using the Visual Language of Comics to Alter
  Sensations in Augmented Reality}. In \bibinfo{booktitle}{\emph{Proceedings of
  the CHI Conference on Human Factors in Computing Systems}} (Honolulu, HI,
  USA) \emph{(\bibinfo{series}{CHI '24})}. \bibinfo{publisher}{Association for
  Computing Machinery}, \bibinfo{address}{New York, NY, USA}, Article
  \bibinfo{articleno}{603}, \bibinfo{numpages}{17}~pages.
\newblock
\showISBNx{9798400703300}
\urldef\tempurl%
\url{https://doi.org/10.1145/3613904.3642351}
\showDOI{\tempurl}


\bibitem[de~Lacerda~Pataca et~al\mbox{.}(2024)]%
        {10.1145/3613904.3642258}
\bibfield{author}{\bibinfo{person}{Calu\~{a} de Lacerda~Pataca},
  \bibinfo{person}{Saad Hassan}, \bibinfo{person}{Nathan Tinker},
  \bibinfo{person}{Roshan~Lalintha Peiris}, {and} \bibinfo{person}{Matt
  Huenerfauth}.} \bibinfo{year}{2024}\natexlab{}.
\newblock \showarticletitle{Caption Royale: Exploring the Design Space of
  Affective Captions from the Perspective of Deaf and Hard-of-Hearing
  Individuals}. In \bibinfo{booktitle}{\emph{Proceedings of the CHI Conference
  on Human Factors in Computing Systems}} (Honolulu, HI, USA)
  \emph{(\bibinfo{series}{CHI '24})}. \bibinfo{publisher}{Association for
  Computing Machinery}, \bibinfo{address}{New York, NY, USA}, Article
  \bibinfo{articleno}{899}, \bibinfo{numpages}{17}~pages.
\newblock
\showISBNx{9798400703300}
\urldef\tempurl%
\url{https://doi.org/10.1145/3613904.3642258}
\showDOI{\tempurl}


\bibitem[de~Lacerda~Pataca et~al\mbox{.}(2023)]%
        {10.1145/3544548.3581511}
\bibfield{author}{\bibinfo{person}{Calu\~{a} de Lacerda~Pataca},
  \bibinfo{person}{Matthew Watkins}, \bibinfo{person}{Roshan Peiris},
  \bibinfo{person}{Sooyeon Lee}, {and} \bibinfo{person}{Matt Huenerfauth}.}
  \bibinfo{year}{2023}\natexlab{}.
\newblock \showarticletitle{Visualization of Speech Prosody and Emotion in
  Captions: Accessibility for Deaf and Hard-of-Hearing Users}. In
  \bibinfo{booktitle}{\emph{Proceedings of the 2023 CHI Conference on Human
  Factors in Computing Systems}} (Hamburg, Germany) \emph{(\bibinfo{series}{CHI
  '23})}. \bibinfo{publisher}{Association for Computing Machinery},
  \bibinfo{address}{New York, NY, USA}, Article \bibinfo{articleno}{831},
  \bibinfo{numpages}{15}~pages.
\newblock
\showISBNx{9781450394215}
\urldef\tempurl%
\url{https://doi.org/10.1145/3544548.3581511}
\showDOI{\tempurl}


\bibitem[faster-whisper({[n.\,d.]})]%
        {fasterwhisper}
faster-whisper \bibinfo{year}{[n.\,d.]}\natexlab{}.
\newblock
\newblock
\newblock
\shownote{\url{https://github.com/SYSTRAN/faster-whisper}}.


\bibitem[Kim et~al\mbox{.}(2023)]%
        {10.1145/3544548.3581130}
\bibfield{author}{\bibinfo{person}{JooYeong Kim}, \bibinfo{person}{SooYeon
  Ahn}, {and} \bibinfo{person}{Jin-Hyuk Hong}.}
  \bibinfo{year}{2023}\natexlab{}.
\newblock \showarticletitle{Visible Nuances: A Caption System to Visualize
  Paralinguistic Speech Cues for Deaf and Hard-of-Hearing Individuals}
  \emph{(\bibinfo{series}{CHI '23})}. \bibinfo{publisher}{Association for
  Computing Machinery}, \bibinfo{address}{New York, NY, USA}, Article
  \bibinfo{articleno}{54}, \bibinfo{numpages}{15}~pages.
\newblock
\showISBNx{9781450394215}
\urldef\tempurl%
\url{https://doi.org/10.1145/3544548.3581130}
\showDOI{\tempurl}


\bibitem[Kudo(2005)]%
        {Kudo2005MeCabY}
\bibfield{author}{\bibinfo{person}{Takumitsu Kudo}.}
  \bibinfo{year}{2005}\natexlab{}.
\newblock \showarticletitle{MeCab : Yet Another Part-of-Speech and
  Morphological Analyzer}.
\newblock
\urldef\tempurl%
\url{https://api.semanticscholar.org/CorpusID:61584143}
\showURL{%
\tempurl}


\bibitem[May et~al\mbox{.}(2023)]%
        {10.1145/3597638.3608398}
\bibfield{author}{\bibinfo{person}{Lloyd May}, \bibinfo{person}{So~Yeon Park},
  {and} \bibinfo{person}{Jonathan Berger}.} \bibinfo{year}{2023}\natexlab{}.
\newblock \showarticletitle{Enhancing Non-Speech Information Communicated in
  Closed Captioning Through Critical Design}. In
  \bibinfo{booktitle}{\emph{Proceedings of the 25th International ACM SIGACCESS
  Conference on Computers and Accessibility}} (New York, NY, USA)
  \emph{(\bibinfo{series}{ASSETS '23})}. \bibinfo{publisher}{Association for
  Computing Machinery}, \bibinfo{address}{New York, NY, USA}, Article
  \bibinfo{articleno}{16}, \bibinfo{numpages}{14}~pages.
\newblock
\showISBNx{9798400702204}
\urldef\tempurl%
\url{https://doi.org/10.1145/3597638.3608398}
\showDOI{\tempurl}


\bibitem[McDonnell et~al\mbox{.}(2024)]%
        {10.1145/3613904.3642177}
\bibfield{author}{\bibinfo{person}{Emma~J McDonnell}, \bibinfo{person}{Tessa
  Eagle}, \bibinfo{person}{Pitch Sinlapanuntakul}, \bibinfo{person}{Soo~Hyun
  Moon}, \bibinfo{person}{Kathryn~E. Ringland}, \bibinfo{person}{Jon~E.
  Froehlich}, {and} \bibinfo{person}{Leah Findlater}.}
  \bibinfo{year}{2024}\natexlab{}.
\newblock \showarticletitle{“Caption It in an Accessible Way That Is Also
  Enjoyable”: Characterizing User-Driven Captioning Practices on TikTok}. In
  \bibinfo{booktitle}{\emph{Proceedings of the CHI Conference on Human Factors
  in Computing Systems}} (Honolulu, HI, USA) \emph{(\bibinfo{series}{CHI
  '24})}. \bibinfo{publisher}{Association for Computing Machinery},
  \bibinfo{address}{New York, NY, USA}, Article \bibinfo{articleno}{492},
  \bibinfo{numpages}{16}~pages.
\newblock
\showISBNx{9798400703300}
\urldef\tempurl%
\url{https://doi.org/10.1145/3613904.3642177}
\showDOI{\tempurl}


\bibitem[Radford et~al\mbox{.}(2022)]%
        {radford2022robustspeechrecognitionlargescale}
\bibfield{author}{\bibinfo{person}{Alec Radford}, \bibinfo{person}{Jong~Wook
  Kim}, \bibinfo{person}{Tao Xu}, \bibinfo{person}{Greg Brockman},
  \bibinfo{person}{Christine McLeavey}, {and} \bibinfo{person}{Ilya
  Sutskever}.} \bibinfo{year}{2022}\natexlab{}.
\newblock \bibinfo{title}{Robust Speech Recognition via Large-Scale Weak
  Supervision}.
\newblock
\newblock
\showeprint[arxiv]{2212.04356}~[eess.AS]
\urldef\tempurl%
\url{https://arxiv.org/abs/2212.04356}
\showURL{%
\tempurl}


\bibitem[whisper-timestamped({[n.\,d.]})]%
        {whispertimestamped}
whisper-timestamped \bibinfo{year}{[n.\,d.]}\natexlab{}.
\newblock
\newblock
\newblock
\shownote{\url{https://github.com/linto-ai/whisper-timestamped}}.


\end{thebibliography}

\appendix

\end{document}